\documentclass[12pt,preprint]{aastex}

\shorttitle{Sloan Bright Arcs Survey}
\shortauthors{Kubo et al.}

\begin{document}


\title{The Sloan Bright Arcs Survey : Discovery of Seven New Strongly Lensed Galaxies from $\rm{z}=0.66-2.94$}

\author{Jeffrey M. Kubo\altaffilmark{1,2}, Sahar S. Allam\altaffilmark{1,2}, Emily Drabek\altaffilmark{2}, Huan Lin\altaffilmark{2}, Douglas Tucker\altaffilmark{2}, Elizabeth J. Buckley-Geer\altaffilmark{2}, H. Thomas Diehl\altaffilmark{2}, Marcelle Soares-Santos\altaffilmark{2,6}, Jiangang Hao\altaffilmark{2}, Matthew Wiesner\altaffilmark{4}, Anderson West\altaffilmark{5},  Donna Kubik\altaffilmark{2}, James Annis\altaffilmark{2},  Joshua A. Frieman \altaffilmark{2,3}}
\altaffiltext{1}{Equal first authors}
\altaffiltext{2}{Center for Particle Astrophysics, Fermi National Accelerator Laboratory, Batavia, IL 60510, USA}
\altaffiltext{3}{Kavli Institute of Cosmological Physics and Department of Astronomy and Astrophysics, University of Chicago, Chicago, IL 60637}
\altaffiltext{4}{Department of Physics, Northern Illinois University, DeKalb, IL 60115 USA}
\altaffiltext{5}{The Illinois Math and Science Academy, Aurora, IL 60506, USA}
\altaffiltext{6}{Instituto de Astronomia, Geofisica e Ciencias Atmosfericas, Universidade de Sao Paulo, Brazil}



\begin{abstract}
We report the discovery of seven new, very bright gravitational lens systems from our ongoing gravitational lens search, the Sloan Bright Arcs Survey (SBAS).  Two of the systems are confirmed to have high source redshifts $z=2.19$ and $z=2.94$.  Three other systems lie at intermediate redshift with $z=1.33,1.82,1.93$ and two systems are at low redshift $z=0.66,0.86$.  The lensed source galaxies in all of these systems are bright, with $i$-band magnitudes ranging from $19.73-22.06$.  We present the spectrum of each of the source galaxies in these systems along with estimates of the Einstein radius for each system.  The foreground lens in most systems is identified by a red sequence based cluster finder as a galaxy group; one system is identified as a moderately rich cluster. In total the SBAS has now discovered 19 strong lens systems in the SDSS imaging data, 8 of which are among the highest surface brightness $z\simeq2-3$ galaxies known.   
\end{abstract}





\section{Introduction}

Strong gravitational lenses allow for the detailed study of distant background galaxies through the magnification provided by lensing.  Models of these systems also provide interesting constraints on the underlying foreground lens mass distribution which includes the distribution of dark matter.  Beginning with the serendipitous discovery of the 8'oclock arc \citep{allam07} we have initiated the Sloan Bright Arcs Survey (SBAS), a systematic search of the Sloan Digital Sky Survey (SDSS) imaging database to search for candidate strong lensed systems.  The first confirmed system from our systematic search was dubbed ``the Clone'' \citep{lin09}, which was followed by the discovery of six other systems described in \citet{kubo09}.  More recently we reported the discovery of four high redshift $z=2$ systems in \citet{diehl09}.  To date we have discovered a dozen strong lens systems by mining the existing SDSS imaging data.

In this Letter we report on a set of seven new systems in the SDSS imaging data which we have confirmed to be bona-fide strong lens systems.  These systems were confirmed via spectroscopy using the Apache Point Observatory (APO) 3.5 m telescope in New Mexico or the Mayall 4 m telescope at Kitt Peak National Observatory.
This letter is organized as follows: candidate selection and follow-up spectroscopy are described in Section 2. In Section 3 we report the details of each system including preliminary mass models and in Section 4 we summarize our results.  Throughout the Letter we use a flat cosmology with $\Omega_M =  0.3$, $\Omega_{\Lambda}=0.7$, and $H_0=100h$~km~s$^{-1}$~Mpc$^{-1}$.






\section{Data}
\label{sec:data}
\subsection{Lens Search}
\label{sec:search}
We searched for candidate strong lens systems in the $8000 \rm{deg^{2}}$ of SDSS imaging data.  Specifically we performed two systematic searches of the SDSS Data Release Five \citep{adelman07} and Data Release Six \citep{adelman08}.  Our first candidate list was generated using the SDSS Catalog Archive Server (CAS) database.  Two separate queries were performed in which we searched for blue objects around a catalog of luminous red galaxies (LRGs; \citep{eisenstein01}) and a catalog of brightest cluster galaxies (BCGs) generated using the maxBCG optical cluster finder algorithm \citep{hansen05}.  Systems were then separated into groups depending on the number of blue objects, $n$, around each LRG or BCG.  Each group of candidates was then visually examined by four different inspectors and flagged if they exhibited an arc-like morphology.   The most promising systems were then chosen for follow-up.  Our query for $n\ge3$ objects produced a list of $1081$ candidates and one confirmed system discovered in this search is presented in Section $\ref{sec:sample}$.

Our second list is generated by performing a search against a catalog of merging galaxies generated using the method outlined in \citet{allam04}.  A merging galaxy pair is defined here as two galaxies in the range $16.0<g<21.0$ which are separated by less than the sum of their respective Pertrosian radii \citep{stoughton02}.  We ran this algorithm on imaging data from the SDSS DR6 and visually examined and classified the resulting catalog.  A total of 5739 candidates were flagged by the algorithm and upon visual inspection a list of 2761 objects were flagged as possible strong lens candidates.  In Section $\ref{sec:sample}$ we report on six systems that have been followed-up from this list which we have confirmed to be strongly lensed galaxies.

\subsection{APO Spectroscopy}

Candidate systems were followed up with the Dual Imaging Spectrograph III (DIS) on the Apache Point Observatory 3.5 m telescope.  DIS is a medium dispersion double spectrograph that has separated red and blue channels.  The standard B400/R300 grating setup was used, which covers a spectral range of $3600-9600$\AA.  The dispersion is 1.83\AA \ per pixel in the blue and 2.31\AA \ per pixel in the red.  We targeted each of our seven arcs with a typical exposure time of $3\times900$s.  We used a $1.5\arcsec$ or $2.0\arcsec$ slit oriented along the brightest segment of the arc in order to maximize the signal-to-noise ratio.  The data were reduced using standard IRAF tasks which we described previously in \citet{lin09} and \citet{diehl09}.  Redshift measurements were done using the {\tt xcsao} task in the IRAF external package {\tt RVSAO} \citep{kurtz98} by cross-correlation \citep{tonry79} against galaxy templates.  For the templates, we used the composite Lyman break galaxy (LBG) template of \citet{shapley03} to cross-correlate against our higher-redshift source galaxies, and we used the galaxy templates from the SDSS DR2 template set\footnote{http://www.sdss.org/dr7/algorithms/spectemplates/index.html} (specifically template numbers 23-28, spanning the range of early- to late-type galaxies) for our lens galaxies and our lower-redshift source galaxies.  

\subsection{Kitt Peak Spectroscopy}

We also followed-up one system, $\rm{SDSS\ J0952+3434}$, with the RC Spectrograph (RCSpec), a low to moderate resolution spectrograph on the Mayall 4 m telescope at Kitt Peak National Observatory.  We observed this system in long slit mode (with slit width of $1.8\arcsec$) and used the KPC-10A grating and the WG-345 order block filter to cover a wavelength range of about $3500-7500$\AA \ (with a dispersion 2.75\AA \ per pixel).  The total exposure time was $4860$s ($3\times900$s and $3\times720$s). Data for this system were also reduced with standard IRAF tasks.

\section{Lens Sample}
\label{sec:sample}
\begin{figure}
\epsscale{1.0}
\plotone{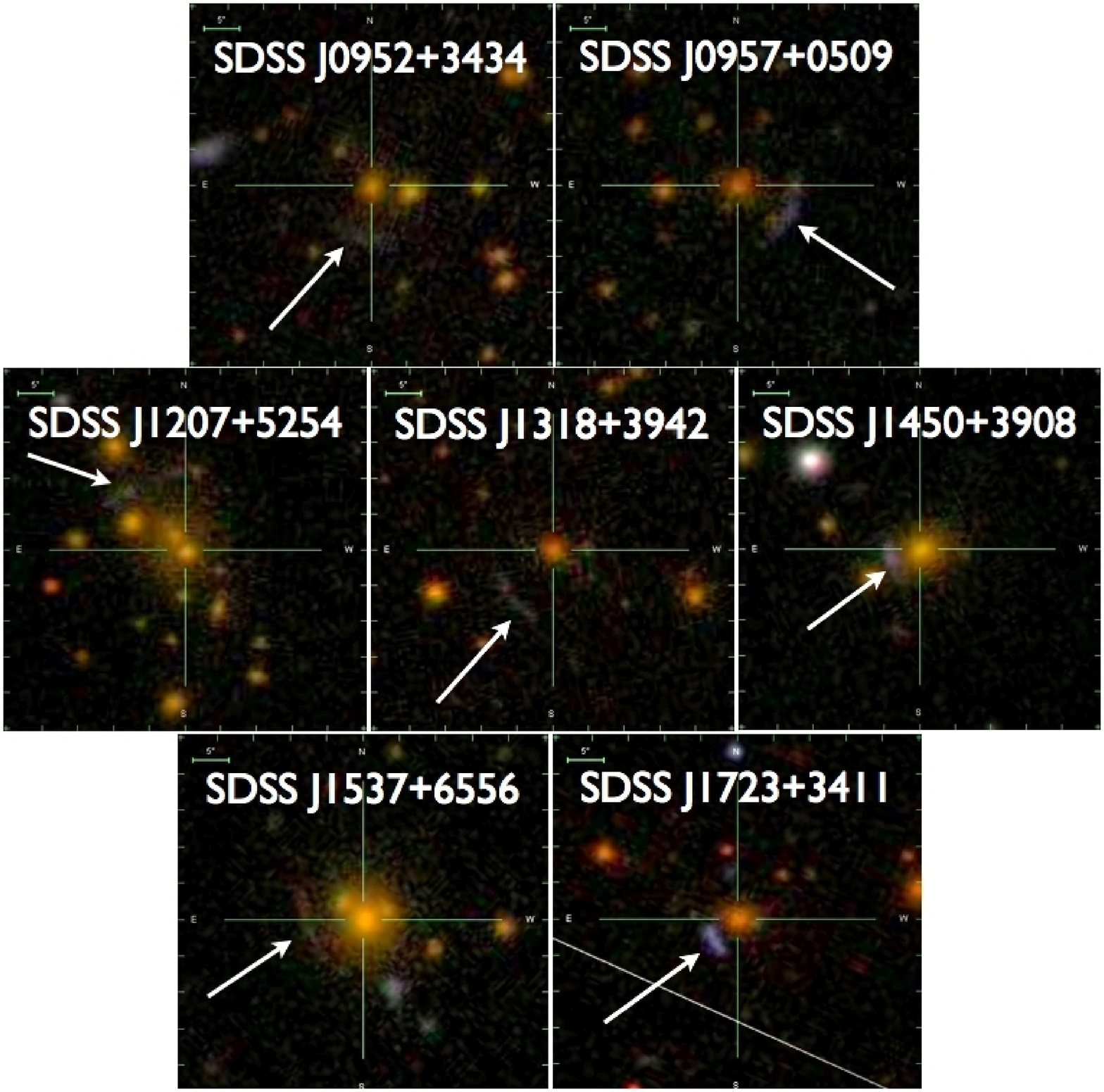}
\caption{Mosaic of seven new strong lens systems discovered in the SDSS.  Images are taken from the SDSS database. Systems are centered on the brighest LRG in each system.  In each image North is up, East is to the left.}
\label{fig:arcs1}
\end{figure}

The seven new strong lens systems are shown in Figure \ref{fig:arcs1}.  Each system has a dominant LRG and bluish arc located within an angular separation of $4.4\arcsec-9.4\arcsec$ of the central LRG.  To characterize the environment of each system we matched the lens coordinates against the gmBCG galaxy cluster catalog \citep{hao10}.  The gmBCG algorithm \citet{hao09} is a red sequence based cluster finding algorithm that calculates a gmBCG richness ($N_{\rm{200}}^{\rm{lens}}$) for each cluster, defined as the number of galaxies within a radius $r_{200}$ centered on the BCG.  The radius $r_{200}$ is the radius in which the mass density falls to 200 times the critical density of the universe \citep{navarro97}.  Six out of seven systems are matched to clusters in the gmBCG catalog, and each of the central LRGs is identified by the gmBCG algorithm as the BCG and also the central galaxy in each group.  One of our lens systems ($\rm{SDSS\ J0952+3434}$) is associated with a moderately rich cluster with $N_{\rm{200}}^{\rm{lens}}=44$, while five other matching systems are associated with lower `group-scale' objects which lie in the richness range $7<N_{\rm{200}}^{\rm{lens}}<11$.  The remaining system ($\rm{SDSS\ J1537+6556}$) did not have a corresponding gmBCG richness measurement (see Section $\ref{sec:arc1}$). 

\begin{figure}
\epsscale{1.0}
\plottwo{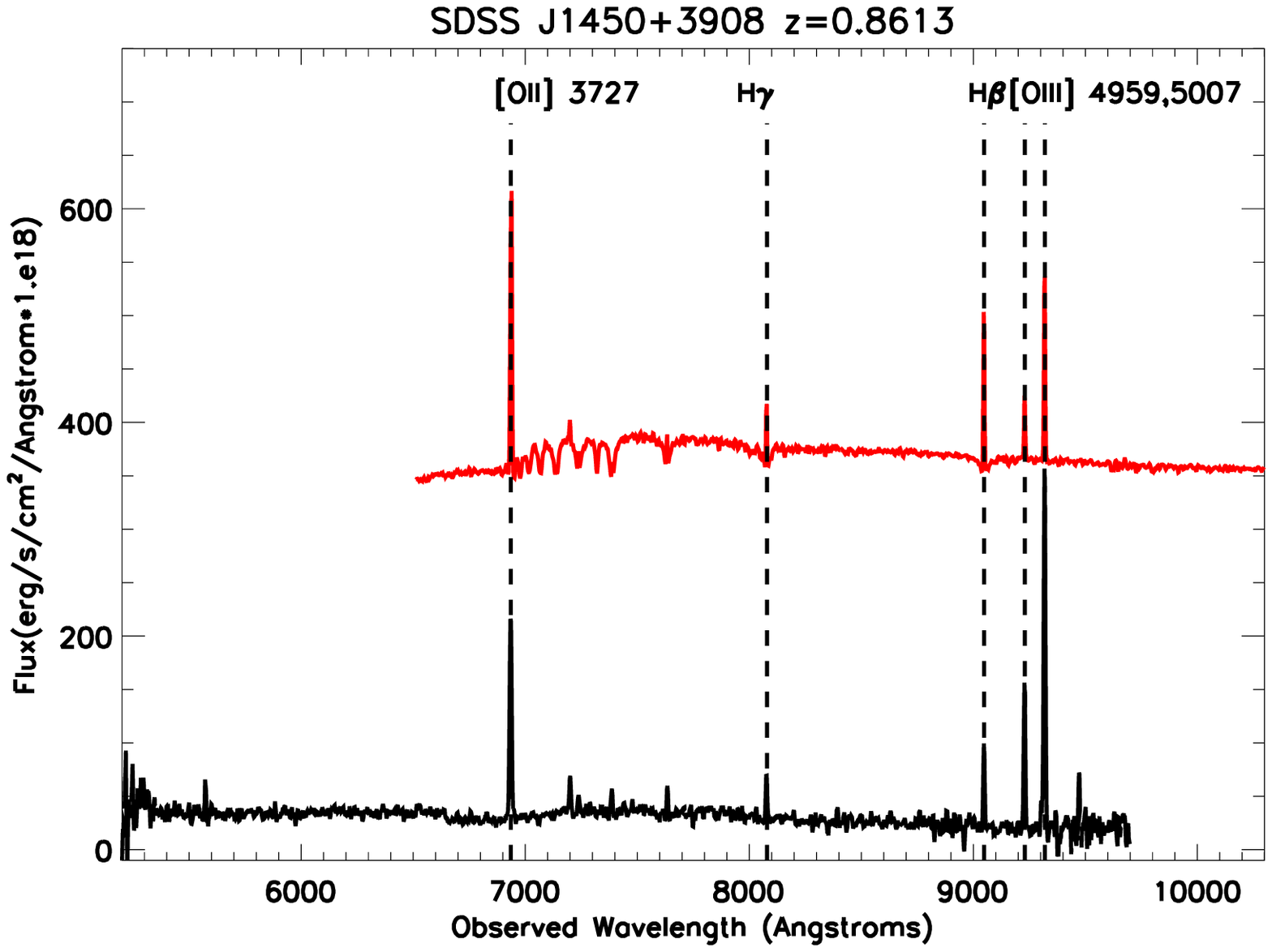}{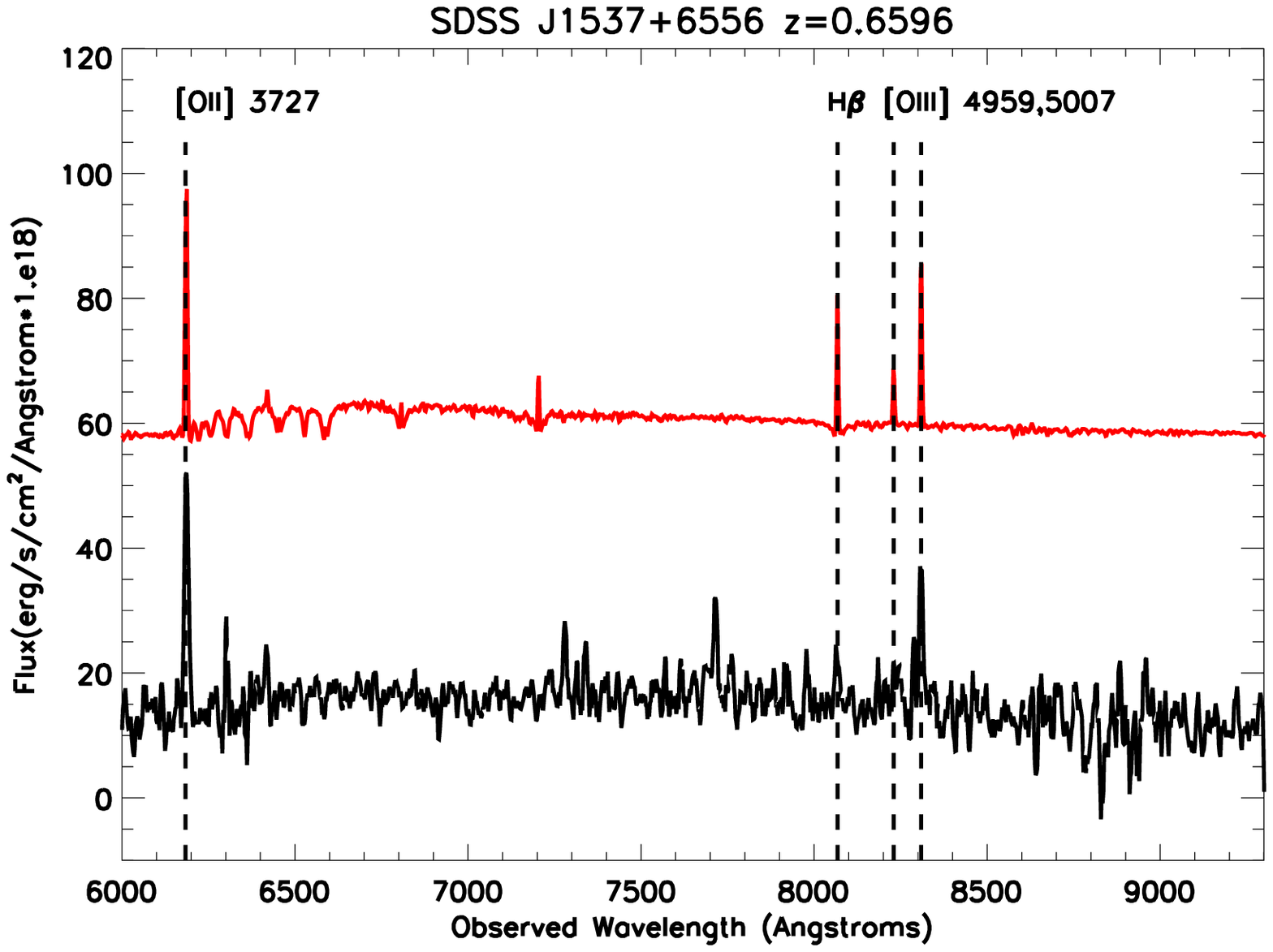}
\caption{DIS spectra of the arcs in two systems (shown in black) : (Left) $\rm{SDSS\ J1450+3908}$ has a redshift of $z=0.8613$; and (Right) $\rm{SDSS\ J1537+6556}$ is at $z=0.6596$.  The DIS spectra are flux-calibrated to $f_\lambda$ units (erg/s/cm$^2$/\AA), but have also been multiplied by $10^{18}$ for plotting purposes.  The red spectrum in each panel is the SDSS emission-line galaxy template used for cross-correlation redshift measurements for these two arcs.  The template has been redshifted to match the arc in each case, and also arbitrarily rescaled and vertically offset for plotting purposes.  The spectra of the arcs are discussed further in $\S \ref{sec:sample}$.}
\label{fig:spectra2}
\end{figure}

Because of the relatively poor image quality of the SDSS images (seeing $\sim1.3\arcsec)$, detailed models of our systems are not possible.  Instead we adopt a simple model and describe each system with a singular isothermal sphere \citep{narayan96}.  The resulting velocity dispersions fall in the range $501-707$ $\rm{km s^{-1}}$, which correspond to enclosed masses between $2.5\times 10^{12} M_{\odot}$ and $10.5\times10^{12} M_{\odot}$.  Magnitudes of the arc in each system are given in Table 1 and are in SDSS model magnitudes unless otherwise stated.  In some cases the arc is split into multiple knots by the SDSS image deblender and in these cases the reported arc magnitude is the total sum of the individual knots.  For one system $(\rm{SDSS\ J1537+6556})$ the arc is not detected in the SDSS photometry and we independently measure and report the isophotal magnitude with SExtractor \citep{bertin96}.  Below we briefly describe each system and present spectra for each arc shown in black in Figures \ref{fig:spectra1} and \ref{fig:spectra2}.  For comparison the composite LBG template\footnote{Because none of our high-redshift source galaxies show Ly $\alpha$ in emission, we have interpolated over the strong Ly $\alpha$ emission line in the LBG template for both cross-correlation and plotting purposes.  We apply this interpolation as it improves the significance of the cross-correlation peak when Ly $\alpha$ is present in the spectrum.} or the SDSS galaxy template is shown in red, shifted to the same redshift of each source galaxy.  A summary of all arc parameters is given in Table \ref{tab:parameters}.

\subsection{$\rm{SDSS\ J0952+3434}$}

\begin{figure}
\plottwo{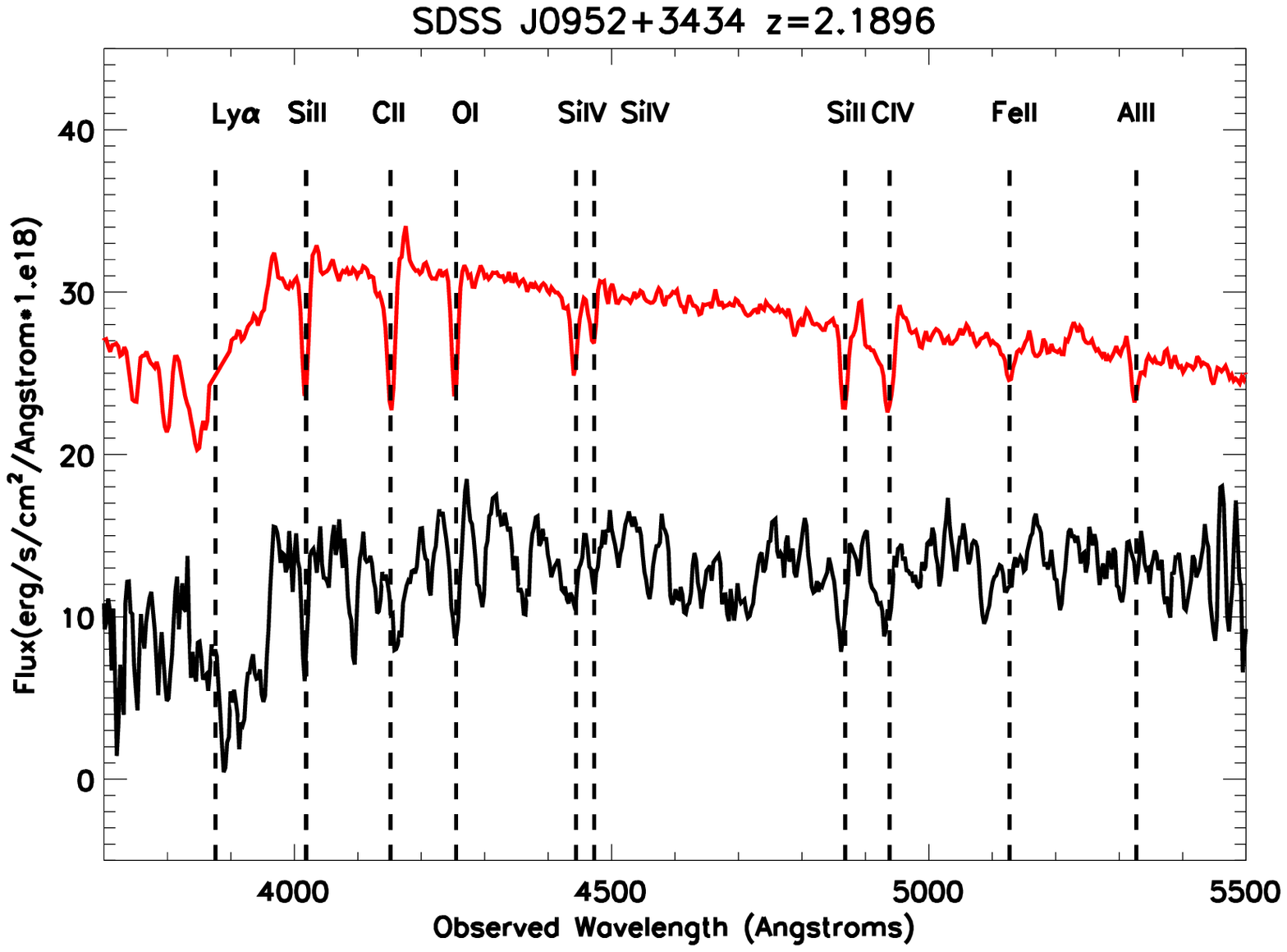}{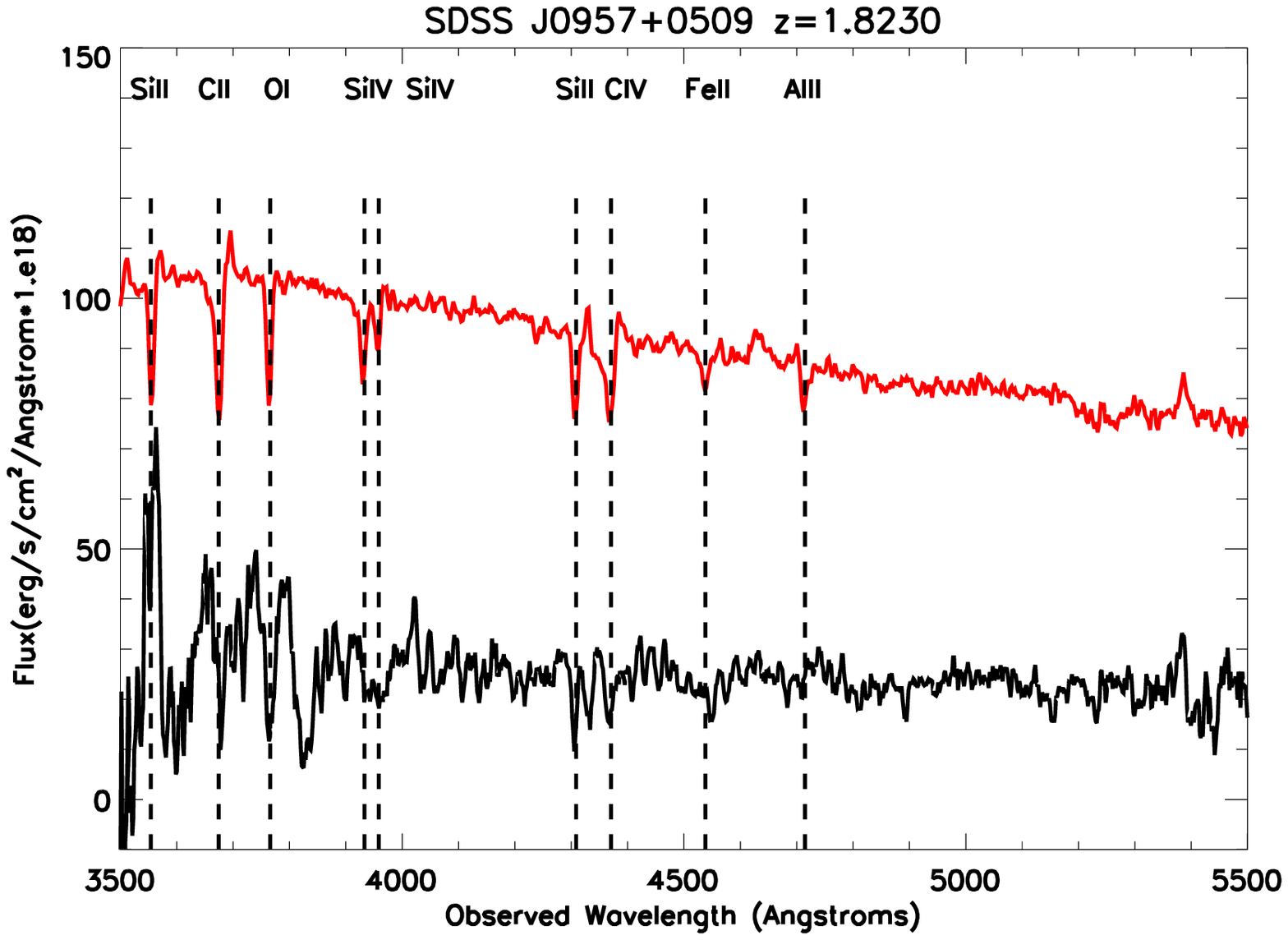}\\
\plottwo{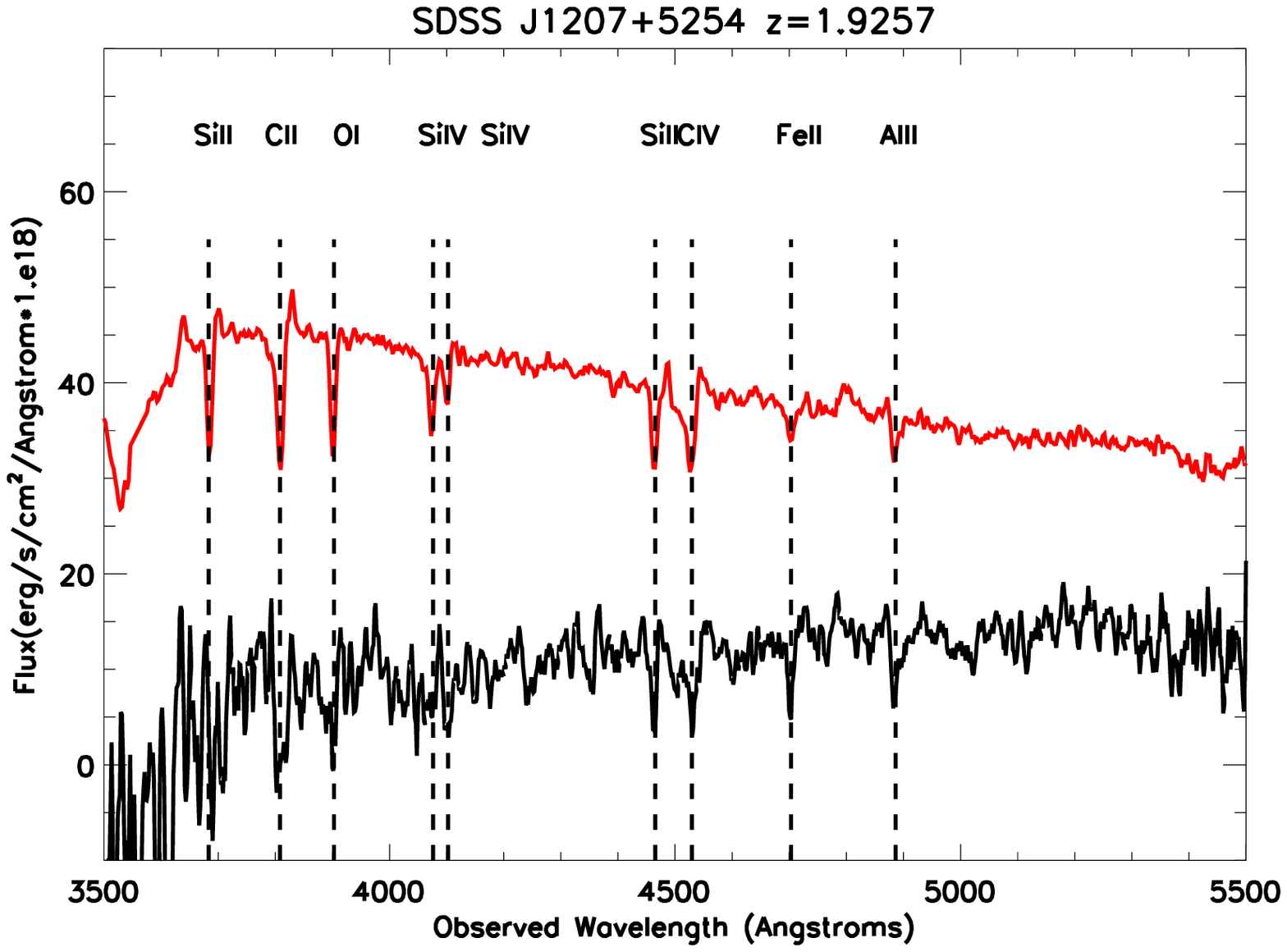}{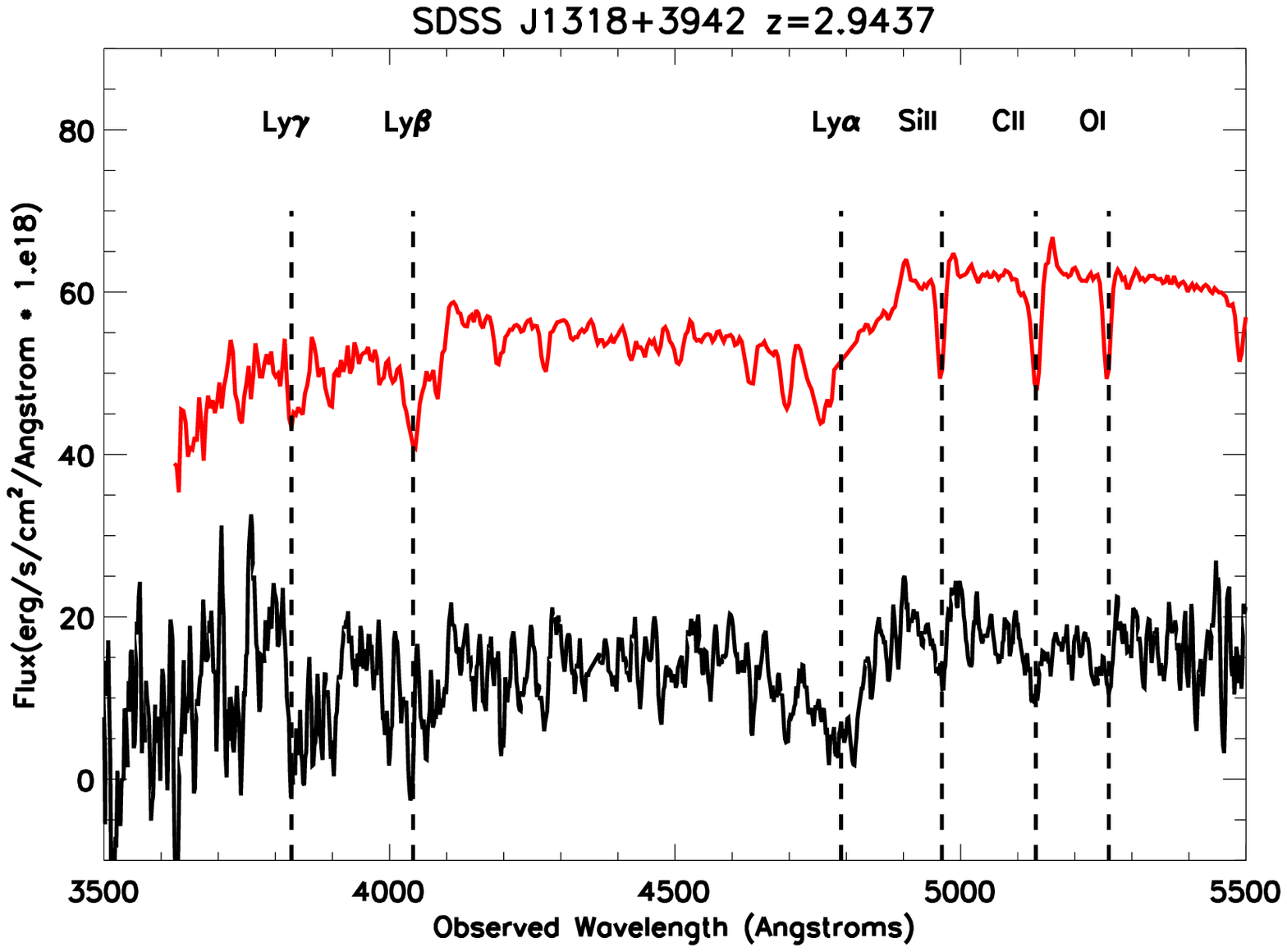}
\caption{DIS spectra of the arcs in four systems are shown in black:  (Upper Left) $\rm{SDSS\ J0952+3434}$ has a redshift of $z=2.1896$ (coadded DIS and RCSpec data); (Upper Right) $\rm{SDSS\ J0957+0509}$ has $z=1.8230$; (Lower Left) $\rm{SDSS\ J1207+5254}$ is at $z=1.9257$; and (Lower Right) $\rm{SDSS\ J1318+3942}$ is at $z=2.9437$.  The DIS spectra are flux-calibrated to $f_\lambda$ units (erg/s/cm$^2$/\AA), but have also been multiplied by $10^{18}$ for plotting purposes.  The red spectrum in each panel is the composite LBG template \citep{shapley03} redshifted to match the arc in each case, and also arbitrarily rescaled and vertically offset for plotting purposes.  The spectra of the arcs are discussed further in $\S \ref{sec:sample}$.  }
\label{fig:spectra1}
\end{figure}

{$\rm{SDSS\ J0952+3434}$ appeared in our search around BCGs and matches to a moderately rich galaxy cluster in the gmBCG cluster catalog with $N_{\rm{200}}^{\rm{lens}}=44$.  The lens consists of two LRGs with a blue arc located directly to the south.  For the eastern-most LRG ($r=21.64$) we measure the spectroscopic redshift of the LRG to be $z=0.3491\pm0.0001$.  For the western LRG ($r=19.51$) we measure a spectroscopic redshift of $z=0.3598\pm0.0001$.  Both LRGs are also members of the cluster as identified by the gmBCG algorithm.  The blue arc is split by the SDSS imaging deblender into two knots, which combined give a total magnitude of $r=21.07$.  Along the arc we used data from both RCSpec and DIS to produce the coadded spectrum (total exposure time of 7560 s) shown in Figure \ref{fig:spectra1}.  Cross-correlation against the LBG template yields a spectroscopic redshift of $z=2.1896\pm0.0001$ for the arc.  The spectrum plotted shows typical LBG absorption features due to Ly $\alpha$, Si {\small{II}}, O {\small{I}}, C {\small{II}}, Si {\small{IV}}, and C {\small{IV}}.  With a measured Einstein radius of $6.9\arcsec$, this corresponds to an enclosed mass of $(5.6\pm0.5)\times 10^{12}M_{\odot}$.

\subsection{$\rm{SDSS\ J0957+0509}$}

$\rm{SDSS\ J0957+0509}$ appeared in our merging galaxy catalog.   The lens is a part of a small group of galaxies with $N_{\rm{200}}^{\rm{lens}}=7$.  The BCG in this group ($r=18.85$) has a spectroscopic redshift determined from DIS to be $z=0.4469\pm0.0002$ (not shown).  A blue arc to the southwest is split into three knots which give an overall arc magnitude of $r=20.04$.  Along the arc we obtain the spectrum plotted in Figure \ref{fig:spectra1}, which shows typical, though relatively weak, LBG absorption features due to Si, C, and O, and we measure a cross-correlation redshift of $z=1.8230\pm0.0003$.  With a measured Einstein radius of $8.0 \arcsec$ this gives an enclosed mass of $(11.3\pm0.9)\times10^{12}M_{\odot}$.

\subsection{$\rm{SDSS\ J1207+5254}$}

$\rm{SDSS\ J1207+5254}$ was flagged as a candidate in our merging galaxy catalog.  The lens is a group of galaxies with $N_{\rm{200}}^{\rm{lens}}=11$, with 2-3 LRGs interior to the arc.  The brightest LRG ($r=17.53$) has a spectroscopic redshift determined with DIS of $z=0.2717\pm0.0002$ (not shown).  This LRG was detected in the SDSS as one object, but based on the imaging may actually be two distinct LRGs.  A third LRG interior to the arc is directed to the northeast with $r=18.64$.  An extended arc is located to the northeast of the central LRG and is also broken into several knots by the SDSS imaging deblender.  The entire arc has a total magnitude of $r=20.56$.  Along the arc we obtain a spectrum that readily shows LBG absorption lines due to C {\small{II}}, O {\small{I}}, Si {\small{IV}}, Si {\small{II}}, C {\small{IV}}, Fe {\small{II}}, and Al {\small{II}}, and we find a cross-correlation redshift for the arc of $z=1.9257\pm0.0002$ (Figure \ref{fig:spectra1}).  With a measured Einstein radius of $9.4\arcsec$ this corresponds to an enclosed mass of $(8.3\pm0.5)\times10^{12}M_{\odot}$.

\subsection{$\rm{SDSS\ J1318+3942}$}

$\rm{SDSS\ J1318+3942}$ was selected from our merging galaxy catalog.  The lens is a group of galaxies ($N_{\rm{200}}^{\rm{lens}}=10$) with a central LRG ($r=18.82$) that has a spectroscopic redshift from the SDSS database of $z=0.4751\pm0.0002$.  To the southeast is a blue arc which has a total magnitude of $r=20.59$.  With DIS we identify strong Ly $\alpha$ absorption, as well as other features due to Si {\small{II}}, C {\small{II}}, O {\small{I}}, Ly $\beta$, and possibly Ly $\gamma$ (Figure \ref{fig:spectra1}).  Cross-correlation against the LBG template places the arc at a redshift of $z=2.9437\pm0.0003$.  We measure the Einstein radius of the system to be $\theta_{E}=8.5\arcsec$ which gives an enclosed mass of $(10.5\pm0.7)\times10^{12}M_{\odot}$. 

\subsection{$\rm{SDSS\ J1450+3908}$}

$\rm{SDSS\ J1450+3908}$} was also selected from our catalog of merging galaxies.   The foreground lens is a group $(N_{\rm{200}}^{\rm{lens}}=8)$ with an LRG $(r=17.62)$ that has a spectroscopic redshift from the SDSS DR7 database of $z=0.2893\pm0.0002$.  To the East of the LRG is a blue arc which is detected as a single object with $r=22.03$.  With DIS we measure five prominent emission lines [O {\small{II}}] 3727, [O {\small{III}}] 4959, [O {\small{III}}] 5007, $H{\small{\beta}}$, and $H{\small{\gamma}}$ (Figure \ref{fig:spectra2}) which place the arc at a redshift $z=0.8613\pm0.00003$.  We measure the Einstein radius to be $\theta_{E}=4.4\arcsec$, which corresponds to an enclosed mass of $(2.5\pm0.3)\times10^{12}M_{\odot}$.

\subsection{$\rm{SDSS\ J1537+6556}$}
\label{sec:arc1}
$\rm{SDSS\ J1537+6556}$ was selected from our merging galaxy catalog.  The lens consists of a central LRG ($r=16.91$) with two fainter galaxies embedded within the halo of the main LRG, one directed to the northeast, with $r=20.77$, and the other directed to northwest, with $r=19.09$.  A blue extended arc is located to the east of the central LRG.  This lens did not appear in the gmBCG catalog since it is located at the edge of the SDSS survey area.    The SDSS photometric software did not converge on a magnitude for the arc so we have independently measured the arc's isophotal magnitude with SExtractor to be $r=20.91$.  For the central LRG, we measure a spectroscopic redshift for the LRG using DIS to be $z=0.2595\pm0.0001$ (not shown).  We did not obtain redshifts for the two foreground galaxies associated with the central LRG.  In the arc we measure four prominent emission lines: [O {\small{II}}] 3727, $H{\small{\beta}}$, [O {\small{III}}] 4959, and [O {\small{III}}] 5007, as shown in Figure \ref{fig:spectra2}, placing the arc at a redshift of $z=0.6596\pm0.0001$.  With a measured Einstein radius of $\theta_{E}=8.1\arcsec$ the enclosed mass for this system is $(7.6\pm0.6)\times 10^{12}M_{\odot}$. 

\subsection{$\rm{SDSS\ J1723+3411}$}
\begin{figure}
\epsscale{1.0}
\plottwo{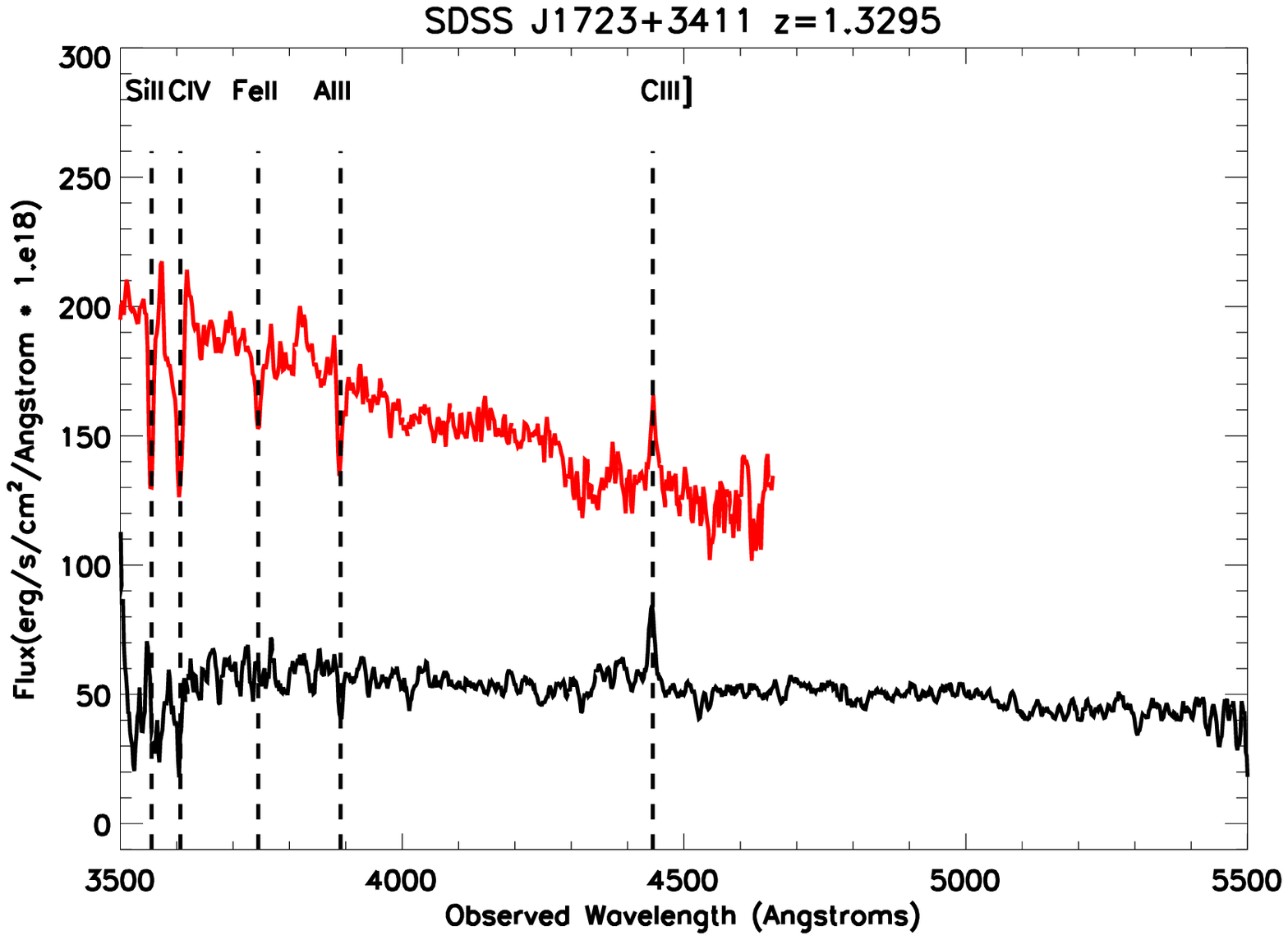}{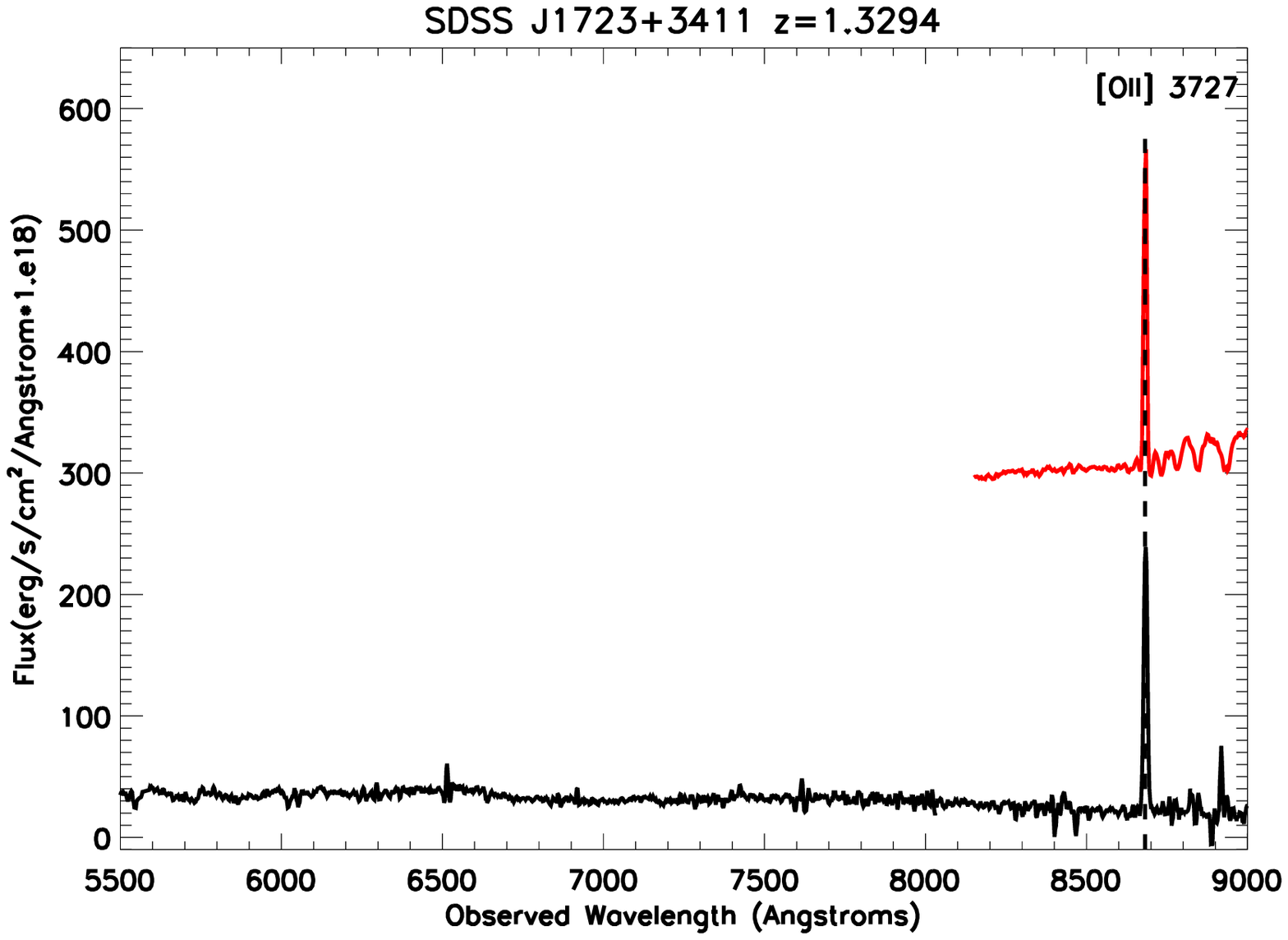}
\caption{DIS spectra of the system $\rm{SDSS\ J1723+3411}$ (shown in black): (Left) the blue portion of the spectrum extending from 3500\AA \ to 5500\AA ; and (Right) the red part of the spectrum extending from 5500\AA \ to 9000\AA.  The DIS spectra are flux-calibrated to $f_\lambda$ units (erg/s/cm$^2$/\AA), but have also been multiplied by $10^{18}$ for plotting purposes.  The red spectrum in each case is the cross-correlation template used for that part of the spectrum, i.e., the composite LBG template in the blue, and an SDSS emission-line galaxy template in the red.  The templates have been redshifted to match the arc in each case, and also arbitrarily rescaled and vertically offset for plotting purposes.  Note the independently determined redshifts from the blue ($z = 1.3295\pm0.0002$) and red ($z =1.3294\pm0.0002$) spectra are consistent with each other.  The spectra are discussed further in $\S \ref{sec:sample}$.}
\label{fig:spectra3}
\end{figure}

$\rm{SDSS\ J1723+3411}$ was selected from our merging galaxy catalog.  The lens is a LRG ($r=18.39$) for which we measure a spectroscopic redshift of $z=0.4435\pm0.0002$ (not shown).  The central LRG is identified by gmBCG as part of a small group with $N_{\rm{200}}^{\rm{lens}}=8$.  A blue arc ($r=21.09$) is located to the southeast and we show both the blue and red parts of the DIS spectrum in Figure \ref{fig:spectra3}.  In the blue part we see an emission line due to C {\small{III}}] 1908, along with possibly several weak LBG absorption features.  In the red part of the spectrum we find a strong emission line due to [O {\small{II}}] 3727.  Independent cross-correlations of the blue spectrum against the composite LBG template, and the red spectrum against an SDSS emission-line galaxy template (number 27), result in consistent redshift measurements, which we then combine to place the arc at a final redshift of $z=1.3294\pm0.0002$.  The measured Einstein radius of the system is $\theta_{E}=4.7\arcsec$, which yields and enclosed mass of $(3.5\pm0.5)\times10^{12}M_{\odot}$.

\section{Summary}
\label{sec:summary}

We have presented seven new strong lens systems from the SBAS, our ongoing search for strongly lensed galaxies in the SDSS imaging data.  Two of these systems are of particular interest since they have source galaxies at high redshift, with $z=2.9437$ and $z=2.1896$.  Because of magnification provided by lensing these are among the brightest galaxies known in this redshift range.  The five remaining systems are source galaxies at intermediate and lower redshifts $z=1.9257,1.8230,1.3294,0.8613,0.6596$.  To date the SBAS has reported a total of 19 strong lens systems, 8 of which have source galaxies at high redshift $z\sim2-3$.  These bright, magnified galaxies are providing important windows into the star formation history and galaxy formation at high redshift.  In upcoming papers we will present detailed models of SBAS systems using high resolution Hubble Space Telescope imaging data as well as other ground-based follow-up data.

\acknowledgments

Fermilab is operated by the Fermi Research Alliance, LLC under Contract No. DE-AC02-07CH11359 with the United States Department of Energy.  These results are based on observations obtained with the Apache Point Observatory 3.5-m telescope, which is owned and operated by the Astrophysical Research Consortium.  Visiting Astronomer, Kitt Peak National Observatory, National Optical Astronomy Observatory, which is operated by the Association of Universities for Research in Astronomy (AURA) under cooperative agreement with the National Science Foundation.  Funding for the SDSS and SDSS-II has been provided by the Alfred P. Sloan Foundation, the Participating Institutions, the National Science Foundation, the U.S. Department of Energy, the National Aeronautics and Space Administration, the Japanese Monbukagakusho, the Max Planck Society, and the Higher Education Funding Council for England. The SDSS Web Site is http://www.sdss.org/.

\clearpage



\begin{deluxetable}{cccccccccc}
\tablecolumns{10}
\rotate 
\tablewidth{0pc}
\tablecaption{Parameters for Each Strong Lens System} 
\tablehead{
\colhead{System} & \colhead{R.A.} & \colhead{Decl.} & \colhead{$z_{l}$} & \colhead{$z_{s}$}& \colhead{$\theta_{E}$\tablenotemark{d}} & \colhead{$\sigma_{v}$} & \colhead{$M(<\theta_{E})$} & \colhead{$(g,r,i)_{\rm{arc}}$}\\
\colhead{} & \colhead{(deg)} & \colhead{(deg)} & \colhead{} & \colhead{}& \colhead{(\arcsec)} & \colhead{(km $\rm{s^{-1}}$)}  & \colhead{($10^{12}\rm{h^{-1}}\rm{M_{\odot}}$)}  & \colhead{}\\
}
\startdata 
$\rm{SDSS\ J0952+3434}$ & 148.16760 &  34.57947 & $0.3491\pm0.0001$\tablenotemark{b} & $2.1896\pm0.0001$\tablenotemark{f}  & $6.9$ & $566\pm12$ & $5.6\pm0.5$ & (21.52,21.07,21.01)\tablenotemark{c}\\
$\rm{SDSS\ J0957+0509}$ & 149.41330 &    5.15887 & $0.4469\pm0.0002$\tablenotemark{b} & $1.8230\pm0.0003$\tablenotemark{b}  & $8.0$ & $651\pm12$ & $9.8\pm0.7$ & (20.55,20.04,19.73)\tablenotemark{c}\\ 
$\rm{SDSS\ J1207+5254}$ & 181.89964 &  52.91645 & $0.2717\pm0.0002$\tablenotemark{b} & $1.9257\pm0.0002$\tablenotemark{b}  & $9.4$  & $644\pm10$ & $8.2\pm0.5$ & (20.81,20.56,20.01)\tablenotemark{c}\\
$\rm{SDSS\ J1318+3942}$ & 199.54796 &  39.70749 & $0.4751\pm0.0002$\tablenotemark{a} & $2.9437\pm0.0003$\tablenotemark{b}  & $8.5$ & $642\pm11$ & $10.5\pm0.7$ & (21.04,20.59,20.24)\tablenotemark{c}\\
$\rm{SDSS\ J1450+3908}$ & 222.62770&  39.13865  & $0.2893\pm0.0002$\tablenotemark{a} & $0.8613\pm0.0003$\tablenotemark{b}  & $4.4$ & $501\pm17$ & $2.5\pm0.3$ & (21.50,22.03,22.06)\tablenotemark{c}\\
$\rm{SDSS\ J1537+6556}$ & 234.30500 &  65.93910 & $0.2595\pm0.0001$\tablenotemark{b} & $0.6596\pm0.0001$\tablenotemark{b}  & $8.1$ & $707\pm13$ & $8.3\pm0.6$ & (22.46,20.91,20.19)\tablenotemark{e}\\
$\rm{SDSS\ J1723+3411}$ & 260.90067 &  34.19946 & $0.4435\pm0.0002$\tablenotemark{b} & $1.3294\pm0.0002$\tablenotemark{b} & $4.7$ & $530\pm17$ & $3.8\pm0.5$ & (20.77,20.99,21.09)\tablenotemark{c}\\
\enddata 
\tablenotetext{a}{Spectroscopic redshift from the SDSS database.}
\tablenotetext{b}{Spectroscopic redshift determined using DIS on the APO 3.5m.}
\tablenotetext{c}{Galaxy model magnitude \citep{stoughton02} from the SDSS database.}
\tablenotetext{d}{Errors on manual fits to $\theta_{E}$ are estimated to be $0.3\arcsec$.}
\tablenotetext{e}{Isophotal magntiudes measured using Sextractor.}
\tablenotetext{f}{Spectroscopic redshift determined using data from DIS and RCSpec.}
 
\label{tab:parameters}
\end{deluxetable}








\clearpage



\clearpage



\clearpage


\end{document}